\documentclass[final]{svjour2}
\usepackage{graphicx}
\usepackage{rotating}
\usepackage{amssymb}
\usepackage{mathptmx}
\usepackage[numbers]{natbib}
\makeatletter
\journalname{Journal of Low Temperature Physics}

\bibpunct{}{}{,}{s}{}{,}

\begin{document}

\newcommand{\hdblarrow}{H\makebox[0.9ex][l]{$\downdownarrows$}-}
\title{Shear modulus in viscoelastic solid $^4$He}

\author{Jung-Jung Su$^{1,2}$ \and Matthias J. Graf$^1$ \and Alexander V. Balatsky $^{1,2}$}

\institute{$^1$Theoretical Division, Los Alamos National Laboratory, Los Alamos, New Mexico 87545, USA \\
$^2$Center for Integrated Nanotechnologies, Los Alamos National Laboratory, Los Alamos, New Mexico 87545, USA}

\date{\today}

\maketitle

\begin{abstract}
The complex shear modulus of solid $^4$He exhibits an anomaly in the same temperature region where torsion oscillators show a change in period. We propose that the observed stiffening of the shear modulus with decreasing temperature can be well described by the response of glassy components inside of solid $^4$He. Since glass is an anelastic material, we utilize the viscoelastic approach to describe its dynamics. The viscoelastic component possesses an increasing relaxation as temperature decreases. The response functions thus derived are identical to those obtained for a glassy, time-delayed restoring back-action. By generalizing the viscoelastic equations for stress and strain to a multiphase system of constituents,
composed of patches with different damping and relaxation properties, we predict that the maximum change of the magnitude of the shear modulus and the maximum height of the dissipation peak are independent of an applied external frequency. The same response expressions allow us to calculate the temperature dependence
of the shear modulus' amplitude and dissipation. Finally, we demonstrate that a Vogel-Fulcher-Tammann (VFT) relaxation time is in agreement with available experimental data.
\end{abstract}


\section{Introduction}
The low-temperature anomaly of solid helium reported for torsional oscillator (TO) by Kim and Chan\cite{Chan04, Chan05} is often regarded as evidence for supersolidity. In addition, the many experiments that followed confirmed that defects in solid $^4$He are important to produce an anomaly. 
Direct experimental evidence for a true phase transition into a supersolid state, on the other hand, remains inconclusive.  To date no definitive sign of Bose-Einstein condensation (BEC) has been seen in measurements of the mass flow \cite{Beamish06,Sasaki06,Balibar08}, the melting curve \cite{Todoshchenko07}, and the lattice structure.\cite{Burns08,Blackburn07}

Controversies exist in interpreting the anomalies, since defects can display their own dynamics and contribute to observables in the parameter range where supersolidity is expected. 
A detailed analysis on the behavior of defects is thus needed to identify the existence of a supersolid state.  
We propose a theoretical framework based on a possible glass component\cite{Balatsky07,Nussinov07} to capture the dynamics of defects in solid $^4$He\cite{Andreev09, glassbook}. This glassy component is suggested to be the cause for the TO and thermodynamic anomalies reported so far. Further, it is consistent with reported signatures of long equilibration times, hysteresis, and a strong dependence on growth history. 
We demonstrated in previous work\cite{Balatsky07,Nussinov07,Graf08,Graf09,Su10} that the freezing out of defect dynamics can account for the anomalies in thermodynamic and  mechanical experiments. The mechanical experiments like torsion oscillator and shear modulus were treated by introducing a back-action term with time delay. 

Based on the same glass hypothesis, we use the viscoelastic approach in this paper to describe the latest shear modulus experimental results. The usage of viscoelastic approach is justified since glass is a viscoelastic material. In fact, our approach here is similar to the model proposed by Yoo and Dorsey\cite{YooDorsey09} for the TO experiments.
We propose the presence of a distribution of viscous components embedded in an otherwise elastic solid. The model we introduce is also known as the generalized Maxwell model. It leads to the same shear modulus expression as the one previously derived within the back-action formulation\cite{Su10_2}. 
This is sensible since the back-action describes how the glass, which is an anelastic material, responds to the external shear stress.  
In addition, we observe that $\omega \tau$ is the scaling parameter of the response functions and we discuss the consequences. Finally, we compare our calculations with available measurements of 
the shear modulus.\cite{Beamish07, Beamish09, Beamish10}

\section{Viscoelastic Model}

\begin{figure}
\begin{center}
\includegraphics[width=.7\linewidth,angle=0,keepaspectratio]{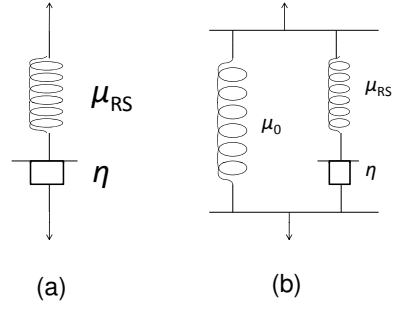}
\end{center}
\vskip-1.cm
\caption{Sketch of a Debye relaxor. (a) A single Debye relaxor element composed of an elastic spring ($\mu_{\rm RS}$) in series with a damped dash-pot ($\eta$). (b) A single Debye relaxor connected in parallel to an elastic spring ($\mu_0$). This viscoelastic element describes a solid with a single relaxation time.  
}\label{fig:singleDebye}
\end{figure}

\begin{figure}
\begin{center}
\includegraphics[width=.8\linewidth,angle=0,keepaspectratio]{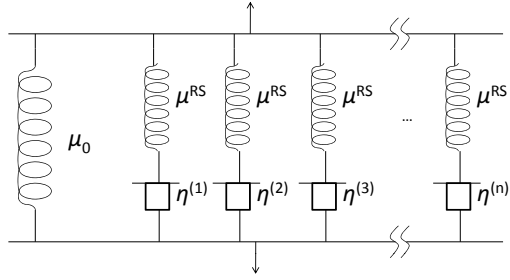}
\end{center}
\vskip-1.5cm
\caption{Sketch of the generalized Maxwell model. The spring $\mu_0$ represents the shear modulus at high temperature.   
}\label{fig:Maxwell}
\end{figure}

The anomalous low-temperature stiffening of the shear modulus can be obtained from a viscoelastic approach. The equivalent circuit model is sketched in Fig.~\ref{fig:singleDebye}.
We assume that the glassy components of the solid give rise to a viscous contribution on top of the dominant elastic behavior of crystalline helium.
Here we describe the coupling between the glassy subcomponent and solid helium by a generalized Maxwell model, shown in Fig.~\ref{fig:Maxwell}.
For pedagogical reasons, we start by considering a single Debye relaxor as plotted in Fig.~\ref{fig:singleDebye}(a). The Debye relaxor is composed of a serial connection of a rigid solid (RS) part, characterized by an elastic shear modulus $\mu_{\rm RS}$, and a Newtonian liquid (NL) part, characterized by a viscosity $\eta$. The RS part describes the ideal elastic solid helium, while NL represents the glassy component, which gives rise to viscous damping. The two parts are connected in series, so that both share the same magnitude of stress, while the net strain is additive. The strain rate equation for both constituents is
\begin{eqnarray}
\dot{\epsilon}= {\dot{\sigma}}/{\mu_{\rm RS}} 
+ {\sigma}/{\eta} ,
\end{eqnarray} 
where $\epsilon$ is the net strain of the Debye relaxor and $\sigma$ is the magnitude of stress shared by the components RS and NL. In order to obtain the above equation, we used the strain relations $\epsilon_{\rm RS}=\sigma_{RS}/\mu_{RS}$ and $\dot{\epsilon}_{\rm NL}=\sigma_{NL}/\eta$. After performing the Fourier transformation we obtain
\begin{eqnarray}
-i \omega \,\epsilon= -i \,\omega \,{\sigma}/{\mu_{\rm RS}} 
+ {\sigma}/{\eta}.
\end{eqnarray}
It follows that the shear modulus of the combined system of such a Debye relaxor (DR) is $\mu_{\rm DR}=\epsilon/\sigma$, 
\begin{eqnarray}
\mu_{DR}(\omega) =\frac{\mu_{\rm RS}}{1+\frac{1}{i \omega \tau_{\rm DR}}} ,
\end{eqnarray}
with relaxation time $\tau_{\rm DR} \equiv \eta/\mu_{\rm RS}$. When the viscoelastic material exhibits a single dominating relaxation time, then it is sufficient to consider the whole solid as a parallel connection between the elastic part connected in parallel to a Debye relaxor,
see Fig.~\ref{fig:singleDebye}(b). Since the shear modulus is additive when connected in parallel, the total shear modulus becomes
\begin{eqnarray} \label{mufinal}
\mu(\omega) &\equiv& \mu_0 + \mu_{\rm DR}(\omega) = \tilde\mu_{0} \left[ 
1 -\frac{g}{ 1- i \omega \tau_{\rm DR} } 
\right] ,
\end{eqnarray}   
with $g = \mu_{\rm RS}/\tilde\mu_0$
and  $\tilde\mu_0=\mu_0+\mu_{\rm RS}$ is the dressed elastic shear modulus. 
In the torsion oscillator and shear modulus experiments the dissipation peak is usually broader than that obtained from a single Debye relaxor. Thus we use a distribution of relaxation times attributed to a distribution of viscoelastic components with their own properties. 
To consider the general case, we need to consider a series of Debye relaxors with different relaxation times connected in parallel as 
shown in Fig.~\ref{fig:Maxwell}. 
Therefore the total contribution from the anelastic part of $n$ constituents in series is given by
$\mu_{ae} = \sum_{n} \mu^{(n)}= \mu_{\rm RS}\sum_{n} [1-1/(1-i \omega \tau_{\rm DR}^{(n)})].$
The continuous version of this expression, when assuming a distribution of relaxation times, $P(t)$, is then
\begin{eqnarray} \label{GMG}
\mu_{ae}(\omega) = \mu_{\rm RS} \int_0^{\infty}
dt \ P(t)
\,\left[1- \frac{1}{1- i \omega \tau t} \right] .
\end{eqnarray}
To make progress, we consider a specific form for $P(t)$. We take the Cole-Cole distribution for dielectric glasses\cite{ColeCole1941}:
\begin{eqnarray}
P(t)=\frac{t^{-(1-\alpha)} \sin \alpha \pi  }
{1+t^{2 \alpha}+2 t^{\alpha}\cos \alpha \pi} .
\end{eqnarray}
In this case, the contribution from the anelastic part to the shear modulus is simply 
\begin{eqnarray} \label{muae}
\mu_{ae}(\omega) = \frac{\mu_{\rm RS}}{ 1-(i \omega \, \tau) ^{\alpha} } .
\end{eqnarray}
Finally, the total shear modulus of the system is essentially that of the anelastic component connected in parallel to the elastic component of the solid,
\begin{eqnarray} \label{mufinal}
\mu(\omega) &=& \tilde\mu_{0} \left[ 
1 -\frac{g}{ 1-(i \omega \tau)^{\alpha} } 
\right] .
\end{eqnarray}
Expression (\ref{mufinal}) is identical to 
the one obtained from a linear-response theory in which an overdamped
glass component is treated as a back-action term.\cite{Su10_2}
This is not a coincidence, since the back-action term describes the response of glass to the external stress and glass is a anelastic material.

\section{Relaxation dynamics}

The expression obtained for the shear modulus in the previous section describes the dynamics of the system by one single scaling parameter $x\equiv \omega \tau$. Hence all response quantities described by $x$ are universal and independent of applied frequency\cite{Zohar}. 

We now discuss response properties that follow  directly from Eq.~(\ref{mufinal}). The experimental observables are the amplitude of the shear modulus, $|\mu|$, and the phase delay between the input and read-out signal, $\phi \equiv  {\rm arg} \, (\mu)$; $\phi$ measures the dissipation of the system, which is related to the inverse of the quality factor $Q^{-1} \equiv \tan \phi$. Defining $\Delta \mu$ as the change in shear modulus amplitude between $x=0$ and $x\to\infty$ one has
\begin{eqnarray}
\frac{\Delta \mu}{\tilde\mu_0} \equiv 
|\mu(x\to\infty)-\mu(x=0)| / \tilde\mu_0
= g, 
\end{eqnarray}
where $g$ measures the strength of the back-action as well as the number of constituents. At finite frequency $\omega>0$, $x \rightarrow \infty$ corresponds to the  low temperature limit, whereas $x \rightarrow 0$ is the high temperature limit.  Thus the applied frequency defines the meaning of ``high'' and ``low'' temperature regime.

Finally, we can estimate the height of the dissipation peak. The dissipation peak is centered at $x=1$ and vanishes at $x=0$ and $x\to\infty$. Hence the peak height is
\begin{eqnarray}  
\Delta \phi \equiv \phi(x=1) - \phi(x=0)
= {\rm arg} \left(1-\frac{g}{1-i^{\alpha}} \right) .
\end{eqnarray}
For $g \ll 1$,  after some straightforward algebra, it reduces to
\begin{eqnarray}
\Delta \phi = \frac{2g}{2-g} \ \cot \left( {\alpha \pi}/{4} \right).
\end{eqnarray}
For $1 < \alpha \le 2 $, this simplifies the expression further to 
\begin{eqnarray} 
\Delta \phi \approx g \cot \left( {\alpha \pi}/{4} \right)
\approx (1-\alpha/2) \left(\frac{\Delta \mu}{\tilde\mu_0} \right) ,
\end{eqnarray} 
where $\Delta \phi$ is in units of radians. The peak height $\Delta \phi$ depends only on the model parameters $\alpha$ and $g$. Note that both $\Delta \mu$ and $\Delta \phi$ are {\it independent of frequency}.

Another important outcome of having a single scaling parameter $x$ in the problem is that the universal Cole-Cole plot is independent of applied frequency.    
We will elaborate on this important behavior and the associated consequences in the next section.

\section{Results}

\begin{figure}
\begin{center}
\includegraphics[width=.75\linewidth,angle=0,keepaspectratio]{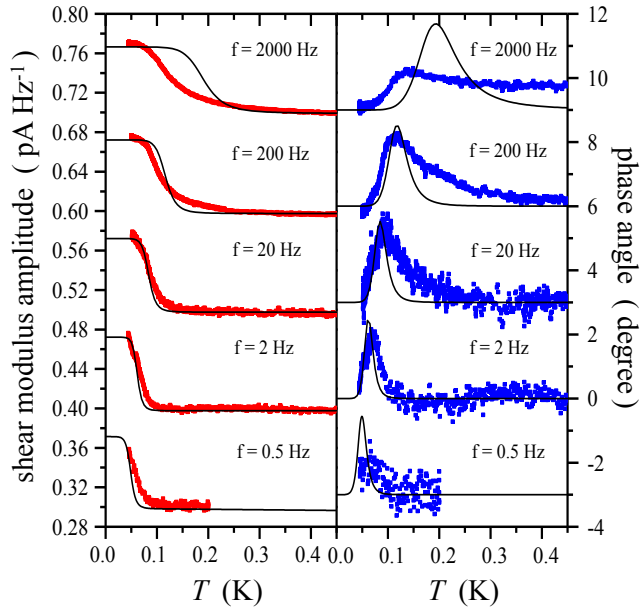}
\end{center}
\vskip-.5cm
\caption{(Color online) 
Experimental data and theoretical calculations of the shear modulus vs.\ temperature assuming a VFT relaxation time. The red and blue squares are the experimental data for the amplitude and dissipation of shear modulus. The black-solid lines show the theoretical calculation. The theoretical calculations use the set of parameters $\alpha=1.31$, $g=1.44\times 10^{-1}, \tilde\mu_0=0.47$ pA Hz$^{-1}$, $\tau_0=50.0$ ns, $\Delta=1.92$K,  and $T_0=-69.3$ mK. Notice a negative $T_0$ means that there is no true phase transition occurring at finite temperatures, probably because of strong quantum fluctuations of helium atoms. The shear modulus amplitude and the phase angle plotted are shifted by 0.1 pA Hz$^{-1}$ and 3 degree with respect to the 2 Hz data.
}\label{fig:SM_VFTnew}
\end{figure}

We compare our theoretical calculations with the experimental data obtained by Day et al\cite{Beamish10} at applied frequencies of 2000 Hz, 200 Hz, 20 Hz, 2 Hz and 0.5 Hz. In particular we like to show the calculation utilizing Vogel-Fulcher-Tamman (VFT) type of relaxation:
\begin{eqnarray}
\tau(T) =
\left\{
\begin{array}{ll}
\tau_0 \,e^{\Delta/(T-T_0)} & \mbox{ for $T>T_0$} , \\
 \infty  & \mbox{ for $T \le T_0$} .
\end{array}
\right.
\end{eqnarray} 
Here $\tau_0$ is the attempt time, $\Delta$ is the activation energy, and $T_0$ is the bare glass transition temperature. The VFT form represents a thermally activated process of relaxors. 
The results are shown in Fig.~\ref{fig:SM_VFTnew}. The data for 2000 Hz and 0.5 Hz show qualitative differences from the other  three consistent data sets (200 Hz, 20 Hz and 2 Hz)\cite{weirddata}. We use the parameter set that best describe the 200 Hz, 20 Hz and 2 Hz data and plot the outcome from the same parameter set for 2000 Hz and 0.5 Hz just for comparison.

We define the crossover temperature $T_X$ as the temperature where the dissipation peaks. The calculated $T_X$ is slightly higher than in experiment for the 200 Hz datasets,  while slightly below the 20 Hz and 2 Hz dataset. As expected $T_{X}$ decreases with decreasing applied frequency.

Notice that although we show the results for VFT relaxation, it is not the only possible relaxation process that can describe the current published data; power-law or other types of relaxation can give similar level of agreement to the experiment (taking into account the number of parameters and the number of data sets being described). In fact, a phenomenological relaxation formulation based also on an anelastic model was used by Syshchenko et al.\cite{Beamish10}. With a significantly broad logarithmic-normal distribution of activation energies in the relaxation formulation, the main feature is also captured. On the other side, Iwasa proposed a relaxation process\cite{Iwasa10} based entirely on dislocation theory of Granato and L\"ucke\cite{GranatoLuecke56}. As we have discussed intensively in our previous publication\cite{Su10_2}, further experiments at lower frequencies and lower temperature are required to characterize the exact type of relaxation.

\begin{figure}[t]
\begin{center}
\includegraphics[width=.8\linewidth,angle=0,keepaspectratio]{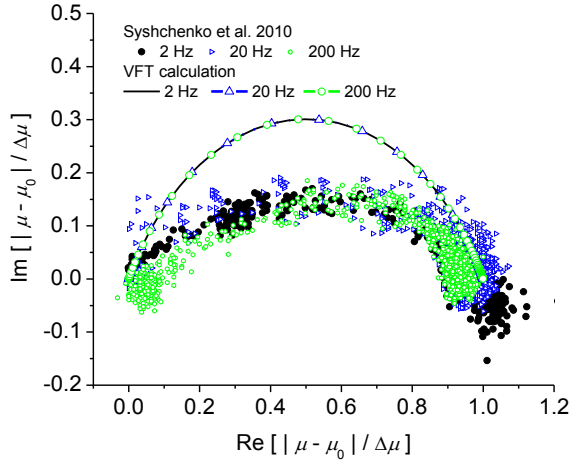}
\end{center}
\vskip-.7cm
\caption{(Color online) 
The Cole-Cole plots for experimental data and for VFT calculation. For given form of $\tau$, all different frequency curves collapse onto one single master curve reflecting that $\omega \tau$ is the only scaling parameter. The Cole-Cole plots show reflection symmetry about Re[$|\mu-\tilde\mu_0|/\Delta \mu$]=0.5, which is a consequence of the Cole-Cole distribution function.         
}\label{fig:CC}
\end{figure}

Finally, we show that the Cole-Cole plots for all three frequencies collapse onto one single curve, see Fig.~\ref{fig:CC}. This is a consequence of $x\equiv\omega \tau$ being the single scaling parameter of the problem, as mentioned in Sec.~3. Note that this feature is universal and should be seen in other dynamic measurements
 besides shear modulus and torsional oscillator experiments.\cite{Hunt09}

\section{Conclusions}
In summary, we have shown that the back-action term of 
a glass contribution can be described by a viscoelastic model through the incorporation of anelastic elements
in the constitutive equations.
When considering the overdamped viscous property of a glass coupling to the elastic helium solid, a generalized Maxwell model leads to 
an identical expression for the complex shear modulus as the glassy back-action term in the linear response theory of elastic deformations.
We have also shown that taking $\omega \tau$ as the scaling parameter of the problem results in a universal scaling behavior of the real and imaginary part of the shear modulus, 
which can be seen in the Cole-Cole plot typical for glasses. Finally, we have shown that our calculations assuming a Vogel-Fulcher-Tammann relaxation time are in accordance with available experimental data.

\begin{acknowledgements}
We acknowledge fruitful discussions with J. Beamish, Z. Nussinov, J. C. Davis, and A. Dorsey. We are especially grateful to Beamish and Syshchenko for sharing their data.
This work was supported by the
 U.S.\ DOE at Los Alamos National Laboratory under contract No.~DE-AC52-06NA25396.
\end{acknowledgements}


\end{document}